\documentclass[10pt,pra,twocolumn,superscriptaddress,showpacs,floatfix]{revtex4}
\usepackage{graphicx}
\usepackage{epstopdf}
\usepackage{amssymb}
\usepackage{times}
\usepackage{amsmath}
\usepackage{amsthm}
\usepackage{amsfonts}%

\begin{document}
\title{Two-way quantum cryptography at different wavelengths}
\author{Christian Weedbrook}
\email{christian.weedbrook@gmail.com}
\affiliation{Department of Physics,
University of Toronto, Toronto, M5S 3G4, Canada}
\author{Carlo Ottaviani}
\email{co@cs.york.ac.uk}
\affiliation{Department of Computer Science, University of York, Deramore Lane, York YO10
5GH, United Kingdom}
\author{Stefano Pirandola}
\email{stefano.pirandola@york.ac.uk}
\affiliation{Department of Computer Science, University of York, Deramore Lane, York YO10
5GH, United Kingdom}
\date{\today}

\begin{abstract}
We study the security of two-way quantum cryptography at different
wavelengths of the electromagnetic spectrum, from the optical
range down to the microwave range. In particular, we consider a
two-way quantum communication protocol where Gaussian-modulated
thermal states are subject to random Gaussian displacements and
finally homodyned. We show how its security threshold (in reverse
reconciliation) is extremely robust with respect to the
preparation noise and able to outperform the security thresholds
of one-way protocols at any wavelength. As a result, improved
security distances are now accessible for implementing quantum key
distribution at the very challenging regime of infrared
frequencies.

\end{abstract}

\pacs{03.67.Dd, 03.67.Hk, 42.50.-p, 89.70.Cf}
\maketitle

\section{Introduction}

Continuous-variable quantum key distribution (QKD)~\cite{Weedbrook2011} allows
the secure distribution of a secret message between two authenticated parties,
Alice and Bob, using nothing more than standard optical equipment, such as
off-the-shelf lasers and homodyne detectors~\cite{Sca09}. Although being
conceived and developed many years after the original discrete-variable
schemes~\cite{Gisin2002}, continuous-variable (CV) QKD has seen many advances and
milestones over the years, both theoretical and
experimental~\cite{Weedbrook2011,Sca09}. More recent advances include, the
security against general attacks in the finite-size
regime~\cite{Leverrier2012}, analysis of practical
imperfections~\cite{Jouguet2012a}, as well as the role of noiseless
amplification~\cite{Blandino2012,Fiurasek2012,Walk2012} and an experimental
demonstration over $80$~km of optical fiber~\cite{Jouguet2012}.

From a practical point of view, there has been recent interest in thermal
QKD~\cite{Fil08,Weedbrook2010,Usenko2010,Weedbrook2012} which essentially
considers the realistic effect of unknown preparation noise on Alice's signal
states. One of the applications of thermal QKD is the ability to generate
secure keys at different wavelengths of the electromagnetic
field~\cite{Weedbrook2010,Weedbrook2012}, from optical range down to the
infrared and microwave regimes. Secure communication at different wavelengths
is ubiquitous in today's communication environment. From the optical
telecommunication wavelength of $1550$~nm down into the GHz microwave regime,
utilized by technologies such as Wi-Fi and cellular phones.

In this paper, we show how to improve the security of thermal QKD at different
wavelengths, using two-way quantum communication, a scheme which is already
known to tolerate higher levels of loss and noise~\cite{S.Pirandola2008}. Variants of the two-way quantum communication protocol also exists~\cite{Sun2012,Zhang2013}. The
idea of using additional preparation noise for two-way quantum communication
was preliminarily investigated in~\cite{Wang2010}. Here they showed the
`fighting noise with noise' effect, an effect first seen in discrete-variable
QKD~\cite{Renner2005} and later in CV one-way
protocols~\cite{Navascues2005,Gar09,Pirandola2009}. Essentially what it shows
is that if extra noise is added in the appropriate way, then the performance
of the protocol can be improved, in terms of secret-key rate and security
threshold~\cite{Weedbrook2011}. In our paper we go further than this initial
analysis by showing the security of the two-way protocol in the presence of
considerably large levels of preparation noise, corresponding to the use of
different communication wavelengths. We show that two-way QKD\ is extremely
robust in reverse reconciliation, such as to beat one-way protocols for any value
of the preparation noise, a feature which allows us to improve the performance
of QKD at the infrared regime.

This paper is structured as follows. In Sec.~II, we introduce one-way thermal
QKD and how this basic setup can be extended to two-way quantum communication.
Then, in Sec~III, we analyze the security of the two-way thermal QKD protocol
by deriving its secret-key rates in direct reconciliation (Sec.~III.A) and
reverse reconciliation (Sec.~III.B). In Section IV, we study the performance
of the two-way thermal protocol at different electromagnetic wavelengths (in
particular, infrared), showing its superiority over one-way thermal QKD.
Finally, Sec.~V is used for the conclusion.

\section{Thermal QKD: From one-way to two-way quantum communication}

\subsection{One-way thermal QKD}

\begin{figure}[th]
\vspace{-0.00cm}
\par
\begin{center}
\includegraphics[width=8cm]{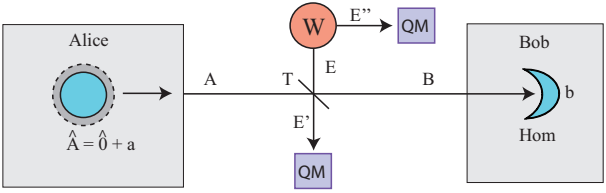}
\end{center}
\par
\vspace{-0.60cm}\caption{One-way thermal QKD protocol. See text for details.}%
\label{pic1}%
\end{figure}

A thermal QKD protocol begins with the sender, Alice, randomly displacing
thermal states in the phase space according to a bivariate Gaussian
distribution. These modulated thermal states are sent to the receiver, Bob,
over an insecure quantum channel which is monitored by the eavesdropper, Eve
(see Fig.~\ref{pic1}). On average, the generic quadrature $\hat{A}$\ of
Alice's input mode $A$ can be written as $\hat{A}=\hat{0}+a$, where the real
number $a$ is the Gaussian encoding variable with variance $V_{a}$, and
$\hat{0}$ is a quadrature operator accounting for the thermal
`preparation noise', with variance
$V_{0}\geq1$ (see Ref.~\cite{Notation1}\ for details on our notation). The
overall variance of Alice's average state is therefore given by $V_{A}%
=V_{0}+V_{a}$.

The variance $V_{0}$ can be broken down as $V_{0}=1+\eta$, where $1$ is the
variance of the vacuum noise, and $\eta\geq0$ is the variance of an extra
noise which is confined in Alice's station and not known to either Eve, Alice
or Bob. At the output of the channel, Bob homodynes the incoming mode $B$,
randomly switching between position and momentum detections (note that an
alternative no-switching protocol based on heterodyne detection can also be
considered~\cite{Weedbrook2004}. In this way, Bob collects an output variable $b$ which is
correlated to Alice's encoding $a$. By making use of a public classical
channel, Bob reveals his basis choices and compares a subset of his data with
Alice in order to estimate the noise in the channel and, therefore, the
maximum information stolen by Eve. If the noise is not too high, compared to a
security threshold, Alice and Bob apply classical post-processing procedures
of error correction and privacy amplification in order to extract a shared secret-key.

In particular, in standard CV-QKD we consider post-processing procedures based
on one-way classical communication, which can be forward from Alice to Bob,
also known as direct reconciliation (DR)~\cite{Grosshans2002}), or backward
from Bob to Alice, also known as reverse reconciliation
(RR)~\cite{Grosshans2003}). In DR, Bob aims to guess Alice's encoding $a$ with
the help of classical communication from her, while in RR it is Alice who aims
to guess Bob's decoding $b$\ with the help of classical communication from
him. Despite Alice and Bob's mutual information not depending on the
direction of the reconciliation, Eve's stolen information does. For this
reason, there are two different secret-key rates defined for the two different
types of reconciliation.

Since the above thermal protocol is a Gaussian protocol~\cite{Weedbrook2011},
its security can be tested against collective Gaussian
attacks~\cite{Nav06,Gar06}, being proven the most powerful attacks allowed by
quantum physics, given a suitable symmetrization of the protocols~\cite{Ren09}%
. Collective Gaussian attacks can be characterized and classified in various
canonical forms~\cite{Pir08}, of which the most important for practical
purposes is represented by the entangling cloner collective
attack~\cite{Gross03}. Such an attack consists in Eve interacting an ancilla
mode $E$ with the signal mode $A$ by means of a beam splitter with
transmission $T\in\lbrack0,1]$. The ancilla mode is part of an
Einstein-Podolski-Rosen (EPR) state $\rho_{EE^{\prime\prime}}$~\cite{Kok2010},
which is an entangled Gaussian state with zero mean and covariance matrix (CM)%
\begin{equation}
\mathbf{V}_{EE^{\prime\prime}}=\left(
\begin{array}
[c]{cc}%
W\mathbf{I} & \sqrt{W^{2}-1}\mathbf{Z}\\
\sqrt{W^{2}-1}\mathbf{Z} & W\mathbf{I}%
\end{array}
\right)  ~,
\end{equation}
where $W\geq1$, $\mathbf{I}=\mathrm{diag}(1,1)$ and $\mathbf{Z}=\mathrm{diag}%
(1,-1)$. Both the kept mode $E^{\prime\prime}$ and the transmitted mode
$E^{\prime}$ are collected in a quantum memory (QM) which is coherently measured at
the end of the protocol.

In a collective entangling-cloner attack, Eve's stolen information can be
over-estimated using the Holevo bound~\cite{Hol73}. In particular, we have to
consider two Holevo quantities $I(E:a)$ or $I(E:b)$ depending if Eve attacks
Alice's variable (DR) or Bob's variable (RR). Subtracting these quantities
from Alice and Bob's classical mutual information $I(a:b)$, we get the two
secret-key rates of the protocol $R^{\blacktriangleright}=I(a:b)-I(E:a)$ and
$R^{\blacktriangleleft}=I(a:b)-I(E:b) $, for DR and RR, respectively. For a
thermal QKD protocol based on modulated thermal states and homodyne detection,
these key rates are functions of the input parameters, namely the variance of
the thermal noise $V_{0}$\ and the variance of the classical signal modulation
$V_{a}$, plus the parameters of the attack, which are the transmission $T$ of
the channel and its thermal variance $W$.

In the typical limit of high modulation ($V_{a}\gg1$), we get the two
analytical expressions for the asymptotic secret-key rates%
\begin{align}
R^{\blacktriangleright}(V_{0},T,W) &  =\frac{1}{2}\log\frac{T\Lambda(W,V_{0}%
)}{(1-T)\Lambda(V_{0},W)}\nonumber\\
&  +h\left[  \sqrt{\frac{W\Lambda(1,WV_{0})}{\Lambda(W,V_{0})}}\right]
-h(W)~,\label{rate1DR}%
\end{align}
and%
\begin{equation}
R^{\blacktriangleleft}(V_{0},T,W)=\frac{1}{2}\log\frac{W}{(1-T)\Lambda
(V_{0},W)}-h(W),\label{rate1RR}%
\end{equation}
where we have used the two functions%
\begin{equation}
\Lambda(x,y):=Tx+(1-T)y~,\label{LambdA}%
\end{equation}
and%
\begin{equation}
h(x):= \Big(\frac{x+1}{2}\Big)\log\Big(\frac{x+1}{2}\Big)-\Big(\frac{x-1}{2}\Big)\log\Big(\frac{x-1}%
{2}\Big)~.\label{h_FUNC}%
\end{equation}
(See Appendix~\ref{app1} for details on their derivation.) By setting these
key rates to zero, we can derive the two security thresholds
$W^{\blacktriangleright}=W^{\blacktriangleright}(V_{0},T)$ and
$W^{\blacktriangleleft}=W^{\blacktriangleleft}(V_{0},T)$ in DR and RR,
respectively. Equivalently, these security thresholds can be given in terms of
tolerable excess noise $N^{\blacktriangleright}=N^{\blacktriangleright}%
(V_{0},T)$ and $N^{\blacktriangleleft}=N^{\blacktriangleleft}(V_{0},T)$, where
$N:=(W-1)(1-T)T^{-1}$.

Note that other types of one-way thermal protocols may be considered, e.g.,
using heterodyne detection instead of homodyne at the
output~\cite{Weedbrook2012}. However, the corresponding security
thresholds, $N^{\blacktriangleright}$ and $N^{\blacktriangleleft}$, are
numerically very similar to those of the one-way protocol just described,
which can be considered as the most representative and practical candidate of
one-way thermal QKD.

\subsection{Two-way thermal QKD}

\begin{figure}[th]
\vspace{-0.4cm}
\par
\begin{center}
\includegraphics[width=8cm]{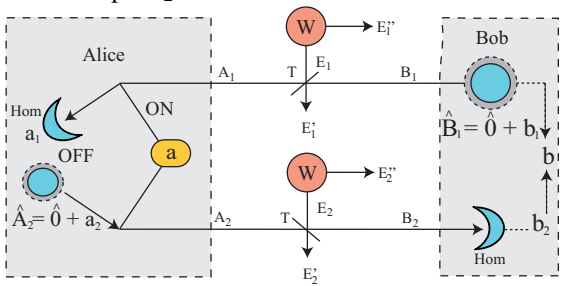}
\end{center}
\par
\vspace{-0.60cm}\caption{Two-way thermal QKD protocol. See text for
explanations. A collective entangling-cloner attack is also shown, where Eve
uses two beam splitters with transmission $T$ and two EPR states with variance
$W$. The output modes $E_{1}^{\prime},E_{1}^{\prime\prime},E_{2}^{\prime
},E_{2}^{\prime\prime}$ are collected in a quantum memory which is coherently
measured at the end of the protocol.}%
\label{fig_2_way}%
\end{figure}

The scenario of Fig.~\ref{pic1} can be readily extended to two-way quantum
communication~\cite{S.Pirandola2008}. As depicted in Fig.~\ref{fig_2_way}, Bob
has an input mode $B_{1}$ where a thermal state with preparation variance
$V_{0}$ is modulated by a bivariate Gaussian distribution with signal variance
$V_{b_{1}}:=\mu$. On average, we have the input quadrature $\hat{B}_{1}%
=\hat{0}+b_{1}$, encoding the Gaussian variable $b_{1}$. In the first quantum
communication through the insecure channel, mode $B_{1}$ is sent to Alice, who
receives the noisy mode $A_{1}$ and randomly switches between two
configurations~\cite{S.Pirandola2008}:

\begin{description}
\item[(i)] ON configuration, where Alice encodes a Gaussian variable $a$ with
variance $V_{a}=\mu$, randomly displacing the quadrature of the incoming mode
$\hat{A}_{1}\rightarrow\hat{A}_{2}=\hat{A}_{1}+a$;

\item[(ii)] OFF configuration, where Alice homodynes the incoming mode $A_{1}
$ with classical output $a_{1}$, and prepares another Gaussian-modulated
thermal state $\hat{A}_{2}=\hat{0}+a_{2}$, with the same preparation and
signal variances as Bob, i.e., $V_{0}$ and $V_{a_{2}}=\mu$.
\end{description}

\noindent In both cases, the processed mode $A_{2}$ is sent back to Bob in the
second quantum communication through the channel. At the output, Bob homodynes
the incoming mode $B_{2}$ with classical output $b_{2}$.

At the end of the double quantum communication, the two parties exploit a
public channel to classically communicate. Alice reveals which configuration,
ON and OFF, was chosen in each round of the protocol, and both the parties
declare which quadratures were detected by their homodyne detectors. After
this stage, Alice and Bob possess a set of correlated variables, which are
$a_{1}\approx b_{1}$ and $a_{2}\approx b_{2}$ in OFF configuration, and
$a\approx b$ in ON configuration, where $b$ is post-processed from $b_{1}$ and
$b_{2}$.

By comparing a small subset of values of these variables, Alice and Bob can
understand which kind of attack has been performed against the two-way
protocol. In particular, the two parties are able to detect the presence of
memory between the first and the second use of the channel. If a memory is
present (two-mode coherent attack), then Alice and Bob use the
OFF\ configuration, extracting a secret-key from $a_{1}\approx b_{1}$ and
$a_{2}\approx b_{2}$. By contrast, if the memory is absent (one-mode
collective attack), then they use the ON\ configuration and they post-process
$a$ and $b$. In other words, the switching between the two configurations is
used as a virtual basis against Eve~\cite{S.Pirandola2008}. Once Alice and Bob
have decided which configuration to use, they post-process their remaining
data using standard one-way algorithms of classical error correction and
privacy amplification, therefore extracting a secret-key in direct or reverse reconciliation.

As discussed in Ref.~\cite{S.Pirandola2008},
%As preliminary discussed in Ref.~\cite{S.Pirandola2008} and
%explicitly proven in~Ref.~\cite{2wayPROOF},
the use of two-mode coherent attacks against the two-way protocol is not
advantageous for Eve. In fact, using the OFF configuration against such
attacks, Alice and Bob can reach security thresholds which are much higher
than those of one-way protocols. Thus, we consider here the analysis of
collective one-mode attacks, in particular, those based on entangling cloners,
which are the most practical benchmark to test CV-QKD. We show that, using the
ON configuration against these attacks, Alice and Bob are able to extract a
secret-key in conditions so noisy that any one-way protocol would fail. In
particular, this happens in reverse reconciliation which turns out to be
extremely robust in the preparation noise, therefore allowing us to improve
the performance of CV-QKD in the very noisy regime of infrared frequencies.

\section{Cryptoanalysis of two-way thermal QKD}

Here we study the security performance of the two-way thermal QKD protocol
against collective entangling-cloner attacks. Adopting the ON\ configuration,
we derive the analytical expressions of the asymptotic secret-key rates (i.e.,
for high modulation $\mu\rightarrow+\infty$), first in DR and then in RR. Such
rates are explicitly plotted in the transmission $T$ for $W=1$ (pure-loss
channel) and studied in terms of the preparation noise $V_{0}$. In the
specific case of RR, we also analyze the behaviour of the security threshold
$W^{\blacktriangleleft}=W^{\blacktriangleleft}(V_{0},T)$ for different values
of $T$ and $V_{0}$, comparing this threshold with that of the corresponding
one-way thermal protocol.

As shown in Fig.~\ref{fig_2_way}, a collective entangling-cloner attack
against the two-way protocol consists of Eve performing two independent and
identical beam-splitter attacks (transmission $T$), one for each use of the
channel. For each beam splitter $i=1$ or $2$, Eve prepares two ancilla modes
$E_{i}$ and $E_{i}^{\prime\prime}$ in an EPR state with variance $W$. Eve
keeps mode $E_{i}^{\prime\prime}$ while injecting the other mode $E_{i}$ into
one port of the beam splitter, leading to the transmitted mode $E_{i}^{\prime
}$. These operations are repeated identically and independently for each
signal mode sent by Bob as well as the return mode sent back to Bob by Alice.
All of Eve's output modes $E_{i}^{\prime}$ and $E_{i}^{\prime\prime}$ are
stored in a quantum memory which is coherently detected at the end of the
two-way protocol. Eve's final measurement is optimized based on Alice and
Bob's classical communication.

For such an attack, Bob's post-processing of his classical variables is just
given by $b=b_{2}-Tb_{1}$. This variable is the optimal linear estimator of
Alice's variable $a$ in the limit of high modulation $\mu\rightarrow+\infty$.
Note that this classical post-processing can be equivalently realized by
constructing a displaced mode $B$ with generic quadrature $\hat{B}=\hat{B}%
_{2}-Tb_{1}$ which is then homodyned by Bob. Despite being useful for the
theoretical analysis of the protocol in RR, such a physical representation is
not practical since it involves the use of a quantum memory to store mode
$B_{2}$ whose displacement $-Tb_{1}$ can only be applied once Bob has
estimated the channel transmission $T$. It is also important to remark that
this equivalent representation is realized by Gaussian operations, so that the
global output state of Bob ($B$-mode) and Eve ($E$-modes) is Gaussian (this is
true for both $b_{1}$ fixed or Gaussian-modulated).

\subsection{Secret-key rate in direct reconciliation}

Let us start our security analysis considering DR~\cite{Notation2}. This type
of reconciliation was shown to be very robust for one-way
protocols~\cite{Weedbrook2010}, which were in principle able to tolerate an
infinite amount of preparation noise and still have a finite secret key,
albeit very small~\cite{Weedbrook2012}. As we show below, such a behavior is
not typical of two-way thermal protocols.

As we know, the secret key rate for DR is given by $R^{\blacktriangleright
}:=I(a:b)-I(a:E)$. The mutual information between Alice and Bob is derived
from the differential Shannon entropy~\cite{Sha48} and is simply given by%
\begin{equation}
I(a:b)=\frac{1}{2}\log_{2}\frac{V_{b}}{V_{b|a}}~,
\end{equation}
where $V_{b}$ is the variance of Bob's post-processed variable $b$, and
$V_{b|a}$ its variance conditioned to Alice's encoding $a$. These variances
are easy to compute once we write the Bogoliubov transformations for the quadratures.

The output mode $B_{2}$\ has generic quadrature%
\begin{equation}
\hat{B}_{2}=T\hat{B}_{1}+\sqrt{T}a+\sqrt{1-T}(\sqrt{T}\hat{E}_{1}+\hat{E}%
_{2})~.
\end{equation}
Subtracting off the input modulation $b_{1}$ (known to only Bob), we get the
processed quadrature $\hat{B}=\hat{B}_{2}-Tb_{1}$ equal to%
\begin{equation}
\hat{B}=T\hat{0}+\sqrt{T}a+\sqrt{1-T}(\sqrt{T}\hat{E}_{1}+\hat{E}_{2})~,
\end{equation}
with variance $V_{B}=T^{2}V_{0}+TV_{a}+(1-T^{2})W$. Since $V_{B}=V_{b}$ and
$V_{a}=\mu$, we get%
\begin{equation}
V_{b}=T^{2}V_{0}+T\mu+(1-T^{2})W~, \label{varianceVb}%
\end{equation}
which gives $V_{b}\rightarrow T\mu$ in the limit of high modulation.

In the same limit, the conditional variance $V_{b|a}$ is given by setting
$\mu=0$ in the previous equation for $V_{b}$, i.e., we have%
\begin{equation}
V_{b|a}=V_{b}|_{\mu=0}=T^{2}V_{0}+(1-T^{2})W~.
\end{equation}
Therefore, the mutual information between Alice and Bob is given by
\begin{align}
I(a:b)&=\frac{1}{2}\log_{2}\frac{T^{2}V_{0}+T\mu+(1-T^{2})W}{T^{2}%
V_{0}+(1-T^{2})W}\nonumber\\
&  \rightarrow\frac{1}{2}\log_{2}\frac{T\mu}{T^{2}V_{0}+(1-T^{2})W}~.
\label{AB_mutual}%
\end{align}

Eve's Holevo information on Alice's encoding variable $a$ is defined as%
\begin{equation}
I(a:E):=S(E)-S(E|a)~,
\end{equation}
where $S(\cdot)$ is the von Neumann entropy of Eve's multimode output state
$\rho_{E}$ (modes $E_{1}^{\prime}E_{1}^{\prime\prime}E_{2}^{\prime}%
E_{2}^{\prime\prime}$) and $S(E|a)$ the entropy of the conditional state
$\rho_{E|a} $ for fixed values of Alice's encoding variable $a$. Since these
states are Gaussian, their entropies can be computed from the symplectic
spectra of their covariance matrices, $\mathbf{V}_{E}$ and $\mathbf{V}_{E|a}$,
respectively~\cite{Weedbrook2011}.

By generalizing the derivation in Ref.~\cite{Pir08} to include the presence of
preparation noise ($V_{0}\geq1$) we get the following expression of Eve's CM
for the Gaussian state $\rho_{E}$ of modes $E_{1}^{\prime}E_{1}^{\prime\prime
}E_{2}^{\prime}E_{2}^{\prime\prime}$%
\begin{equation}
\mathbf{V}_{E}(V_{a},V_{a})=\left(
\begin{array}
[c]{cc|cc}%
\varepsilon\mathbf{I} & \varphi\mathbf{Z} & \chi\mathbf{I} & \mathbf{0}\\
\varphi\mathbf{Z} & W\mathbf{I} & \theta\mathbf{Z} & \mathbf{0}\\\hline
\chi\mathbf{I} & \theta\mathbf{Z} & \mathbf{\Delta}(V_{a},V_{a}) &
\varphi\mathbf{Z}\\
\mathbf{0} & \mathbf{0} & \varphi\mathbf{Z} & W\mathbf{I}%
\end{array}
\right)  , \label{EvemainCM}%
\end{equation}
where $\mathbf{0}:=\mathrm{diag}(0,0)$ and the parameters are defined as
\begin{align}
\varepsilon &  :=(1-T)V_{B_{1}}+TW,\\
\chi &  :=-\sqrt{T}(1-T)(W-V_{B_{1}}),\\
\theta &  :=-(1-T)(W^{2}-1),\\
\gamma &  :=T(1-T)V_{B_{1}}+(1-T+T^{2})W\\
\varphi &  :=\sqrt{T(W^{2}-1)},\\
\mathbf{\Delta}(V_{a},V_{a})  &  :=\gamma\mathbf{I}+(1-T)~\mathrm{diag}%
(V_{a},V_{a})~.
\end{align}
In the previous parameters, we set $V_{B_{1}}=V_{0}+\mu$ and $V_{a}=\mu$, and
we consider the limit of high modulation ($\mu\rightarrow+\infty$). Thus, we
are able to compute the asymptotic symplectic spectrum of the CM\ which is
given by the four eigenvalues $\nu_{1}\rightarrow W$, $\nu_{2}\rightarrow W$
and $\{\nu_{3},\nu_{4}\}$ such that $\nu_{3}\nu_{4}\rightarrow(1-T)^{2}\mu
^{2}$. Using these eigenvalues, we compute the entropy of Eve's state
$\rho_{E}$ which is given by~\cite{Hol99}
\begin{equation}
S(E)=\sum_{k=1}^{4}h(\nu_{k})\rightarrow2h(W)+\log\left(  \frac{e}{2}\right)
^{2}(1-T)^{2}\mu^{2},
\end{equation}
where we have used the $h$-function of Eq.~(\ref{h_FUNC}) and its asymptotic
expansion $h(x)\simeq\log(ex)/2$ for large $x$.

Now we consider the conditional CM\ $\mathbf{V}_{E|a}$ which is given by
$\mathbf{V}_{E}(0,\mu)$~\cite{Pir08} As a result, $\mathbf{V}_{E|a}$ is the
same as $\mathbf{V}_{E}$ except for the adjustment of the variable
$\mathbf{\Delta}(\mu,\mu)\rightarrow\mathbf{\Delta}(0,\mu)$. In the usual
limit ($\mu\rightarrow+\infty$) we compute the conditional spectrum $\bar{\nu
}_{1}\rightarrow1$, $\bar{\nu}_{2}\rightarrow W$ and $\{\bar{\nu}_{3},\bar
{\nu}_{4}\}$ such that $\bar{\nu}_{3}\bar{\nu}_{4}\rightarrow(1-T)\sqrt
{(1-T^{2})W\mu^{3}}$. Such eigenvalues allow us to derive the conditional
entropy $S(E|a)$ and, therefore, to compute Eve's Holevo information%
\begin{equation}
I(a:E)\rightarrow h(W)+\frac{1}{2}\log\frac{(1-T)\mu}{(1+T)W}~. \label{EVEdr}%
\end{equation}
Combining Eqs.~(\ref{AB_mutual}) and~(\ref{EVEdr}), we get following
asymptotic expression for the DR\ secret-key rate%
\begin{align}
R^{\blacktriangleright}(V_{0},T,W)  &  =\frac{1}{2}\log\frac{T(1+T)W}%
{(1-T)(T^{2}V_{0}+(1-T^{2})W)}\nonumber\\
&  -h(W)
\end{align}
\begin{figure}[th]
\vspace{-0.0cm}
\par
\begin{center}
\includegraphics[width=8cm]{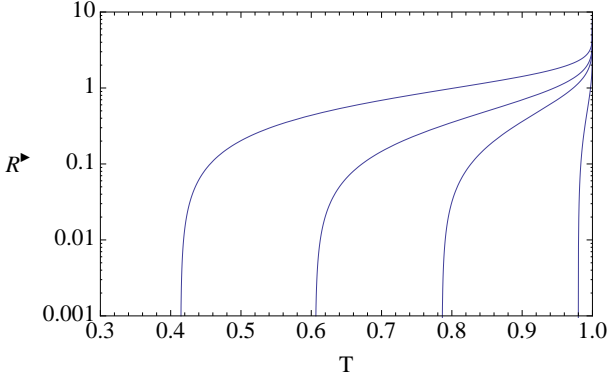}
\end{center}
\par
\vspace{-0.5cm}\caption{Plot of the DR\ secret-key rate of the two-way thermal
protocol for a pure-loss channel ($W=1$) as a function of the channel
transsmissivity $T$, for different values of the preparation noise
$V_{0}=1,5,10$ and $100$ (from left to right).}%
\label{fig_coh_hom}%
\end{figure}

In order to study the performance of the two-way thermal protocol, we plot in
Fig.~\ref{fig_coh_hom} the DR secret-key rate in the presence of a pure-loss
channel (corresponding to an entangling-cloner attack with $W=1$) as a
function of the channel transmission $T$, and for different values of the
preparation noise $V_{0}$. As we can see from the figure, two-way quantum
communication with modulated pure states ($V_{0}=1$) is able to beat the
$3$~dB loss limit (corresponding to the threshold $T=1/2$). However, as the
preparation noise is increased, we see a fairly rapid deterioration in the
security of the protocol. Such a behavior is different from what happens in
DR for the corresponding one-way thermal
protocol~\cite{Weedbrook2010,Weedbrook2012}. In fact, despite one-way thermal
QKD being secure only within the $3$~dB loss limit, such limit is not affected
by the preparation noise $V_{0}$, so that high values of $V_{0}$ are tolerable
in the range $0.5<T<1$ with the DR secret-key rate remaining positive even
if close to zero.

However, contrarily to what happens in DR, we show below that two-way thermal
QKD is very robust to the preparation noise in RR, such that its security
threshold outperforms both the thresholds (in DR and RR) of the one-way
thermal QKD at any value of the preparation noise $V_{0}$. This is the feature
that we will exploit to improve the security at the infrared regime.

\subsection{Secret-key rate in reverse reconciliation}

Let us derive the RR secret-key rate $R^{\blacktriangleleft}:=I(a:b)-I(E:b)$%
~\cite{Notation2}. Here we need to compute Eve's Holevo information on Bob's
processed variable $b$, i.e., $I(E:b)=S(E)-S(E|b)$. From the formula, it is
clear that we need to compute the entropy $S(E|b)$ of Eve's output state
$\rho_{E|b}$ conditioned to Bob's variable $b$. To compute the CM
$\mathbf{V}_{E|b}$ of this state, we first derive the global CM
\begin{equation}
\mathbf{V}_{EB}=\left(
\begin{array}
[c]{cc}%
\mathbf{V}_{E} & \mathbf{D}\\
\mathbf{D}^{T} & V_{b}\mathbf{I}%
\end{array}
\right)  ~,
\end{equation}
describing Eve's modes $E_{1}^{\prime}E_{1}^{\prime\prime}E_{2}^{\prime}%
E_{2}^{\prime\prime}$, with reduced CM\ $\mathbf{V}_{E}$ given in
Eq.~(\ref{EvemainCM}), plus Bob's virtual mode $B$, with reduced CM
$V_{b}\mathbf{I}$, with the variance $V_{b}$ computed in Eq.~(\ref{varianceVb}%
). Then, we apply homodyne detection on mode $B$, which
provides~\cite{Eisert2002,Fiurasek2002,Weedbrook2011,Gae} $\mathbf{V}%
_{E|b}=\mathbf{V}_{E}-(1/V_{b})\mathbf{D}\mathbf{\Pi D}^{T}$, where
$\mathbf{\Pi}:=\mathrm{diag}(1,0,0,0)$. Here the off-diagonal block
$\mathbf{D}$ describes the correlations between Eve's and Bob's modes, and is
given by
\begin{equation}
\mathbf{D}^{T}=\left(
\begin{array}
[c]{cccc}%
\xi_{1}\mathbf{I}, & \phi_{1}\mathbf{Z}, & \xi_{2}\mathbf{I}, & \phi
_{2}\mathbf{Z}%
\end{array}
\right)  ~,
\end{equation}
where
\begin{align}
\xi_{1}  &  =-T\sqrt{1-T}(V_{0}-W),\\
\phi_{1}  &  =\sqrt{T(1-T)}\sqrt{W^{2}-1},\\
\xi_{2}  &  =-\sqrt{T(1-T)}(TV_{0}+V_{a})+TW\sqrt{T(1-T)},\\
\phi_{2}  &  =\sqrt{1-T}\sqrt{W^{2}-1}~.
\end{align}
In the previous formulas, we set $V_{a}=V_{b_{1}}=\mu$ and we consider the
limit of high modulation $\mu\rightarrow+\infty$. In this limit, we derive the
asymptotic expression of the conditional symplectic spectrum $\{\tilde{\nu
}_{1},\tilde{\nu}_{2},\tilde{\nu}_{3},\tilde{\nu}_{4}\}$, which is given by
$\tilde{\nu}_{1}\rightarrow W$,
\begin{equation}
\tilde{\nu}_{2}\rightarrow\sqrt{\frac{W\left(  1+T^{2}V_{0}W+T^{3}%
(1-V_{0}W)\right)  }{T^{2}V_{0}+W+T^{3}(W-V_{0})}}~,
\end{equation}
and%
\begin{equation}
\tilde{\nu}_{3}\tilde{\nu}_{4}\rightarrow\sqrt{\frac{(1-T)^{3}\left(
T^{2}V_{0}+W+T^{3}\left(  W-V_{0}\right)  \right)  \mu^{3}}{T}}~.
\end{equation}
Using this spectrum we compute the conditional entropy $S(E|b)$ and,
therefore, the RR secret-key rate, whose asymptotic expression is equal to
\begin{align}
R^{\blacktriangleleft}(V_{0},T,W)  &  =\frac{1}{2}\log\frac{T^{2}V_{0}%
+W+T^{3}\left(  W-V_{0}\right)  }{\left(  V_{0}T^{2}+\left(  1-T^{2}\right)
W\right)  (1-T)}\nonumber\\
&  +h\left(  \tilde{\nu}_{2}\right)  -h(W)
\end{align}
\begin{figure}[th]
\vspace{-0.3cm}
\par
\begin{center}
\includegraphics[width=8.5cm]{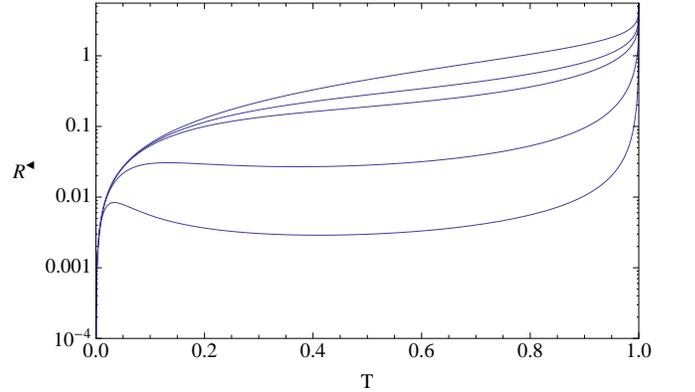}
\end{center}
\par
\vspace{-0.4cm}\caption{Plot of the RR secret-key rate of the two-way thermal
protocol for a pure-loss channel ($W=1$) as a function of the transmissivity
$T$ for different values of the preparation noise $V_{0}=1,5,10,100$ and
$1000$ (from top to bottom). As the preparation noise is increased, the rate
decreases but remains positive for any $T>0$.}%
\label{fig_coh_hom_2way}%
\end{figure}

In Fig.~\ref{fig_coh_hom_2way} we plot the RR secret-key rate
$R^{\blacktriangleleft}$ in the presence of a pure-loss channel ($W=1$) as a
function of the channel tranmissivity $T$ for values of the preparation noise
from $V_{0}=1$ to $V_{0}=10$. As we can see, there is no reduction in the
security of the protocol, in the sense that all the curves originate from the
common threshold $T=0$ for any value of the preparation noise $V_{0} $, even
if the rate is decreasing for increasing $V_{0}$. This is clearly in contrast
to what happened before for DR (with the transmission threshold $T$
approaching $1$ for high values of $V_{0}$).

\begin{figure}[th]
\vspace{-0.3cm}
\par
\begin{center}
\includegraphics[width=7.5cm]{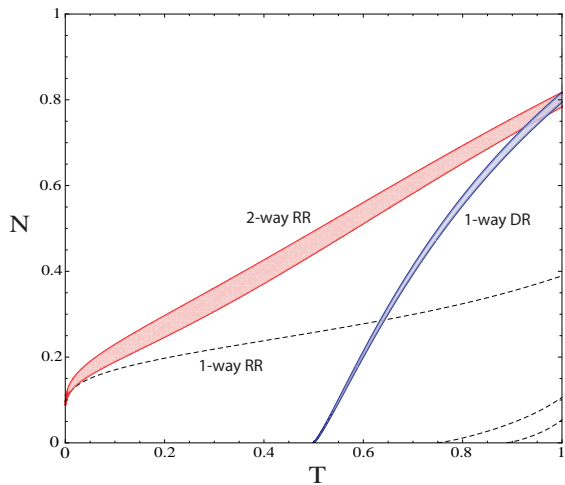}
\end{center}
\par
\vspace{-0.4cm}\caption{(Color online). Two-way thermal protocol in the
presence of an arbitrary entangling-cloner attack ($W\geq1$). We plot the RR
security threshold, expressed as tolerable excess noise $N^{\blacktriangleleft
}$ as a function of the channel transmissivity $T$ for different values of the
preparation noise from $V_{0}=1$ to $10^{6}$ (illustrated by the shaded regions). This threshold is compared with
the DR threshold of the one-way thermal protocol for $V_{0}=1$ to $10^{6}$.
The plot also shows the RR threshold of the one-way protocol (dashed curves)
for $V_{0}=1,5,10$ (from left to right).}%
\label{ThresholdsCOMP}%
\end{figure}

Next, we analyze the security of the two-way thermal protocol against an
arbitrary entangling-cloner attack (with $W\geq1$). In
Fig.~\ref{ThresholdsCOMP}, we plot its RR security threshold, expressed as
tolerable excess noise $N^{\blacktriangleleft}=N^{\blacktriangleleft}%
(V_{0},T)$ as a function of the transmissivity $T$, for a wide range of values
of the preparation noise $V_{0}$. As we can see, $N^{\blacktriangleleft}$ is
very robust with respect to the preparation noise $V_{0}$, with all the
curves, from $V_{0}=1$ up to $V_{0}=10^{6}$, being included in the region
shown in the figure. Thus, despite the RR secret-key rate
$R^{\blacktriangleleft}$ being decreasing for increasing $V_{0}$, it remains
positive up to the excess noise $N^{\blacktriangleleft}$ shown in
Fig.~\ref{ThresholdsCOMP}. (Furthermore, the threshold value
$N^{\blacktriangleleft}$ turns out to be slightly increasing in $V_{0}$, as a
result of the `fighting noise with noise' effect of QKD).

From the same figure, we can see that the two-way thermal protocol outperforms
the one-way thermal protocol in the transmission range $0<T<1$ for any
value of the preparation noise $V_{0}$ (up to $10^{6}$). In fact, the one-way
protocol is not robust in RR (see dashed curves in the figure), and its DR
security threshold $N_{\text{1-way}}^{\blacktriangleright}$ is well below the
two-way RR\ threshold $N_{\text{2-way}}^{\blacktriangleleft}$, apart from a
small overlapping region very close to $T=1$. By exploiting this robustness
and better performance of the two-way thermal protocol, we can improve the
security of CV-QKD at the infrared regime as discussed in the following
section.

\section{Two-way thermal QKD at different wavelengths}

By exploiting its robustness to the preparation noise, we can use the two-way
thermal protocol to improve the security of CV-QKD at longer
wavelengths. Given a bosonic mode with frequency $f$ in a thermal bath with
temperature $t$, it is described by a thermal state with mean number of
photons $\bar{n}=[\exp(hf/k_{B}t)-1]^{-1}$, where $h$ is Planck's constant and
$k_{B}$ is Boltzmann's constant~\cite{Ger05}. This number gives the
noise-variance of the thermal state $V_{0}=2\bar{n}+1$, i.e., the preparation
noise, which is therefore function of the frequency and the temperature, i.e.,
$V_{0}=V_{0}(f,t)$. In our study we consider a fixed value of the temperature
$t=15~$%
%TCIMACRO{\U{b0}}%
%BeginExpansion
${{}^\circ}$%
%EndExpansion
C, so that $V_{0}$ is one-to-one with the frequency $f$.

Eve's attack is a collective entangling-cloner attack (as before) which is
thought to be performed inside a cryostat. The purpose of this is to remove
Eve from the background preparation noise at any wavelength, therefore making
her ancillary modes pure. In order to cover her tracks, Eve uses an entangling
cloner with channel noise equal to the preparation noise, i.e., $W=V_{0}$. For
more information on how to implement an entangling-cloner attack in thermal
QKD, see the details given in \cite{Weedbrook2012}.

\begin{figure}[th]
\begin{center}
\includegraphics[width=8cm]{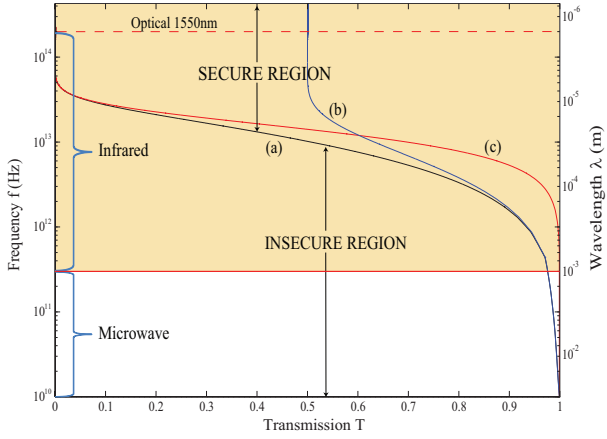}
\end{center}
\caption{Security threshold of the
two-way thermal protocol in RR (a) compared to the security thresholds of the
one-way thermal protocol in DR (b) and RR (c). Thresholds are expressed as minimum
tolerable frequency (or maximum tolerable wavelengths) as a function of the
channel transmission $T$. Note how the two-way threshold is deeper in the
infrared regime. Environmental temperature is $t=15~$${{}^{\circ}}$C.}%
\label{2way_thermal2}%
\end{figure}

Thus, at fixed temperature ($t=15~$%
%TCIMACRO{\U{b0}}%
%BeginExpansion
${{}^\circ}$%
%EndExpansion
C), we can express the RR secret-key rate $R^{\blacktriangleleft}(V_{0},T,W)$
as a function of $f$ and $T$, i.e., $R^{\blacktriangleleft}%
=R^{\blacktriangleleft}(f,T)$. By setting $R^{\blacktriangleleft}=0$, we get
the security threshold $f^{\blacktriangleleft}=f^{\blacktriangleleft}(T)$,
giving the minimum tolerable frequency $f^{\blacktriangleleft}$ which can be
used at any channel transmission $T$ or, equivalently, the maximum tolerable
wavelenght $\lambda^{\blacktriangleleft}=c/f^{\blacktriangleleft}$, with $c$
being the speed of light. The threshold $f^{\blacktriangleleft}(T)$ is plotted
in Fig.~\ref{2way_thermal2} and compared with the thresholds of the one-way
thermal protocol in DR and RR~\cite{Weedbrook2010,Weedbrook2012}. As we can
see from the figure, two-way QKD allows us to use a broader range of
frequencies than one-way QKD. In particular, this happens for $0.2\lesssim
T\lesssim0.8$, where the two-way threshold is well below the other thresholds
in the infrared regime. 

From Fig.~\ref{2way_thermal2}, we see a crossing point between the DR and RR
thresholds of the one-way protocol, for $T\simeq0.6$ and $f\simeq
1.2\times10^{13}~$Hz. This point identifies the maximum gap from the two-way
configuration, which remains secure for channel transmissions as low as
$T\simeq0.4$ at the same crossing frequency $f\simeq1.2\times10^{13}~$Hz. Such
a frequency corresponds to a wavelength of about $\lambda=24~\mu$m, an
infrared region where quantum communication is very demanding, with free-space
losses around 9.7~dB/m under ideal atmospheric conditions (with humidity equal
to $1$mm of water vapor column and temperature of $15$~%
%TCIMACRO{\U{b0}}%
%BeginExpansion
${{}^\circ}$%
%EndExpansion
C~\cite{geminiIR}).

As a result, one-way QKD\ ($T=0.4$) is secure up to a distance of $22$~cm,
while two-way QKD\ ($T=0.6$) remains secure up to $41$~cm. Despite being a
very short distance, this represents an improvement close to $100\%$, which
could be extremely useful in short-range cryptography, e.g., for connecting
close computers through infrared ports or interfacing mobile devices with ATM machines.

Note that the infrared regime is less challenging in the $10~\mu$m window, for
instance at $\lambda=12~\mu$m. At this wavelength, the atmospheric absorption
is dominated by carbon dioxide, methane, and ozone, with an attenuation which
is much smaller (about 0.53 dB/km). In this case, one-way QKD\ is secure up to
$14.6$~Km, while two-way QKD allows the parties to distribute secret-keys up
to $15.8$~Km, corresponding to an $8\%$ improvement in the distance.

\section{Conclusion}

In conclusion, we have shown how the use of two-way quantum communication
enables two parties to improve the security of thermal QKD, where
(inaccessible) preparation noise is added to the signal states. Considering
both types of reconciliation procedures (direct and reverse reconciliation), we have analyzed the
secret-key rates and the security thresholds of a two-way protocol which is
based on Gaussian-modulated thermal states, random Gaussian displacements
and homodyne detections. We have tested its security against collective
Gaussian (entangling-cloner) attacks, showing how the security threshold in reverse reconciliation
is very robust with respect to the preparation noise, and is able to
outperform the security thresholds (in direct and reverse reconciliation) of one-way thermal QKD.

Exploiting such a robustness, we have successfully extended two-way thermal
QKD\ to longer wavelengths, where thermal background naturally provides very
high values of preparation noise. In particular, we have shown the superiority
of two-way quantum communication in the infrared regime, improving the
security distances which can be reached by the use of thermal sources at such frequencies.

\acknowledgments

C.O. thanks Giovanni Di Giuseppe for useful discussion. C.W.
acknowledges support from NSERC. The work of S.P. and C.O. has
been supported by EPSRC under the research grant EP/J00796X/1
(HIPERCOM).

%%%%% begin appendix  %%%%%%%%
\appendix

\section{Asymptotic secret-key rates for one-way thermal QKD\label{app1}}

In this appendix, we provide the main steps for deriving the asymptotic
secret-key rates of Eqs.~(\ref{rate1DR}) and~(\ref{rate1RR}), for the one-way
thermal protocol in DR\ and RR, respectively. These formulas are the explicit
analytical expressions of the key rates already studied in
Refs.~\cite{Weedbrook2010,Weedbrook2012}.

One can easily verify that the CM of Eve's output modes $E^{\prime}%
E^{\prime\prime}$ is given by%
\begin{equation}
\mathbf{V}_{E}=\left(
\begin{array}
[c]{cc}%
W\mathbf{I} & \sqrt{T(W^{2}-1)}\mathbf{Z}\\
\sqrt{T(W^{2}-1)}\mathbf{Z} & TW+(1-T)(V_{a}+V_{0})\mathbf{I}%
\end{array}
\right)  ~. \label{eq:V1wayEVE}%
\end{equation}
For large modulation ($V_{a}\gg1$), we can easily compute its symplectic
spectrum~\cite{Weedbrook2011}, which is given by $\nu_{1}\rightarrow W$ and
$\nu_{2}\rightarrow(1-T)V_{a}$. Then, Eve's entropy is equal to
\begin{equation}
S(E)=h(\nu_{1})+h(\nu_{2})\rightarrow h(W)+\log\frac{e}{2}(1-T)V_{a}~,
\label{eq:Se1wayDR}%
\end{equation}
where $h(\cdot)$ is defined in Eq.~(\ref{h_FUNC}). In the same limit of large
modulation, Eve's conditional CM $\mathbf{V}_{E|a}$ can be computed from
$\mathbf{V}_{E}$ by setting $V_{a}=0$ for one of the two quadratures. One can
then check that $\mathbf{V}_{E|a}$ has the following asymptotic eigenvalues%
\begin{equation}
\bar{\nu}_{1}\rightarrow\sqrt{\frac{W\Lambda(1,WV_{0})}{\Lambda(W,V_{0})}%
},~\bar{\nu}_{2}\rightarrow\sqrt{(1-T)\Lambda(W,V_{0})V_{a}}~,
\end{equation}
where the function $\Lambda(x,y)$ is defined in Eq.~(\ref{LambdA}). As a
result, Eve's conditional entropy is given by%
\begin{equation}
S(E|a)\rightarrow h(\bar{\nu}_{1})+\log\frac{e}{2}\sqrt{(1-T)\Lambda
(W,V_{0})V_{a}}~. \label{eq:Sea1wayDR}%
\end{equation}
Combining Eqs.~(\ref{eq:Se1wayDR}) and~(\ref{eq:Sea1wayDR}), we get Eve's
Holevo information%
\begin{align}
I(E  &  :a)=S(E)-S(E|a)\nonumber\\
&  \rightarrow h(W)-h(\bar{\nu}_{1})+\frac{1}{2}\log\frac{(1-T)V_{a}}%
{\Lambda(W,V_{0})}~. \label{eq:Xe1wayDR}%
\end{align}

Very easily, we can also compute Alice and Bob's mutual information, which is
given by%
\begin{equation}
I(a:b)=\frac{1}{2}\log\left[  1+\frac{TV_{a}}{\Lambda(V_{0},W)}\right]  ~.
\label{eq:Iab1way}%
\end{equation}
Using Eqs.~(\ref{eq:Xe1wayDR}) and~(\ref{eq:Iab1way}), we derive the
asymptotic secret-key rate in DR, equal to Eq.~(\ref{rate1DR}).

Let us now compute the key rate in RR. Note that Eve and Bob's joint CM is
given by%
\begin{equation}
\mathbf{V}_{EB}=\left(
\begin{array}
[c]{cc}%
\mathbf{V}_{E} & \mathbf{C}\\
\mathbf{C}^{T} & \mathbf{B}%
\end{array}
\right)  ~,
\end{equation}
where $\mathbf{B}=[\Lambda(V_{0},W)+TV_{a}]\mathbf{I}$, and%
\begin{equation}
\mathbf{C}=\left(
\begin{array}
[c]{c}%
\sqrt{(1-T)(W^{2}-1)}\mathbf{Z}\\
\sqrt{T(1-T)}[W-(V_{a}+V_{0})]\mathbf{I}%
\end{array}
\right)  ~.
\end{equation}
\newline As a result of Bob's homodyne detection, Eve's conditional CM is
given by $\mathbf{V}_{E|b}=\mathbf{V}_{E}-\mathbf{C}(\boldsymbol{\Pi
}\mathbf{B}\boldsymbol{\Pi})^{-1}\mathbf{C}^{T}$, with $\boldsymbol{\Pi
}:=\mathrm{diag}(1,0)$~\cite{Eisert2002,Fiurasek2002,Weedbrook2011,Gae}. For
large modulation, this CM has asymptotic spectrum $\tilde{\nu}_{1}%
\rightarrow1$, and%
\begin{equation}
\tilde{\nu}_{2}\rightarrow\sqrt{\frac{1-T}{T}WV_{a}}~,
\end{equation}
so that Eve's conditional entropy is given by%
\begin{equation}
S(E|b)\rightarrow\log\frac{e}{2}\sqrt{\frac{1-T}{T}WV_{a}}~.
\end{equation}
Using the expressions for $S(E)$ and $I(a:b)$ previously computed, we can
derive the asymptotic secret-key rate in RR which is given in
Eq.~(\ref{rate1RR}).


\bigskip

\begin{thebibliography}{99}                                                                                               %


\bibitem {Weedbrook2011}C.~Weedbrook, S.~Pirandola, R.~Garc\'ia-Patr\'on,
N.~J.~Cerf, T.~C.~Ralph, J.~H.~Shapiro, and S.~Lloyd, Rev.~Mod.~Phys.
\textbf{84}, 621 (2012).

\bibitem {Sca09}V.~Scarani, H.~Bechmann-Pasquinucci, N.~J.~Cerf, M.~Dusek,
N.~Lutkenhaus, and M.~Peev, Rev.~Mod.~Phys. \textbf{81}, 1301 (2009).

\bibitem {Gisin2002}N.~Gisin, G.~Ribordy, W.~Tittel, and H.~Zbinden, Rev. Mod.
Phys. \textbf{74}, 145 (2002).

\bibitem {Leverrier2012}A.~Leverrier, R.~Garc\'ia-Patr\'on, R.~Renner, and
N.~J.~Cerf, Phys. Rev. Lett. \textbf{110}, 030502 (2013).

\bibitem {Jouguet2012a}P.~Jouguet, S.~Kunz-Jacques, E.~Diamanti, and
A.~Leverrier, Phys. Rev. A \textbf{86}, 032309 (2012).

\bibitem {Fiurasek2012}J.~Fiur\`{a}\v{s}ek and N.~J.~Cerf, Phys. Rev. A \textbf{86}, 060302(R) (2012).

\bibitem {Blandino2012}R.~Blandino, A.~Leverrier, M.~Barbieri, J.~Etesse,
P.~Grangier, and R.~Tualle-Brouri,  Phys. Rev. A \textbf{86}, 012327 (2012).

\bibitem {Walk2012}N.~Walk, T.~Symul, P.~K.~Lam, T.~C.~Ralph, Phys. Rev. A. \textbf{87}, 020303, (2013).

\bibitem {Jouguet2012}P.~Jouguet, S.~Kunz-Jacques, A.~Leverrier, P.~Grangier,
and E.~Diamanti, Nature Photonics \textbf{7}, 378 (2013).

\bibitem {Fil08}R.~Filip, Phys. Rev. A \textbf{77}, 022310 (2008).

\bibitem {Usenko2010}V.~C.~Usenko and R.~Filip, Phys.~Rev.~A \textbf{81},
022318 (2010).

\bibitem {Weedbrook2010}C.~Weedbrook, S.~Pirandola, S.~Lloyd, and T.~C.~Ralph,
Phys. Rev. Lett. \textbf{105}, 110501 (2010).

\bibitem {Weedbrook2012}C.~Weedbrook, S.~Pirandola, and T.~C.~Ralph, Phys.
Rev. A \textbf{86}, 022318 (2012).

\bibitem {S.Pirandola2008}S. Pirandola, S. Mancini, S. Lloyd, and S. L.
Braunstein, Nature Phys. 4, \textbf{726} (2008).

\bibitem{Sun2012} M. Sun, X. Peng, Y. Shen, and H. Guo, Int. J. Quantum Inform.
\textbf{10}, 1250059 (2012).

\bibitem{Zhang2013} Y.~-C.~Zhang, Z.~Li, C.~Weedbrook, S.~Yu, W.~Gu, M.~Sun, X.~Peng, H.~Guo, arXiv:1307.7590 (2013).



\bibitem {Wang2010}M.~Wang and W.~Pan, Phys. Lett. A \textbf{374}, 2434 (2010).

\bibitem {Renner2005}R.~Renner, N.~Gisin, and B.~Kraus, Phys. Rev. A
\textbf{72}, 012332 (2005).

\bibitem {Navascues2005}M.~Navascu\'{e}s and A. Ac\'{\i}n, Phys. Rev. Lett.
\textbf{94}, 020505 (2005).

\bibitem {Gar09} R.~Garc\'ia-Patr\'on
and N.~J.~Cerf, Phys. Rev. Lett. \textbf{102}, 130501 (2009).

\bibitem {Pirandola2009}S.~Pirandola, R.~Garc\'{\i}a-Patr\'{o}n,
S.~L.~Braunstein, and S.~Lloyd, Phys. Rev. Lett. \textbf{102}, 050503 (2009).

\bibitem {Notation1}To simplify the notation, in our paper we work in the
Heisenberg picture showing how the generic quadratures of the modes evolve.
For instance, $\hat{A}$ represents either the position-quadrature $\hat{Q}%
_{A}$\ or the momentum-quadrature $\hat{P}_{A}$ of mode $A$. Gaussian
modulations are intended to be independent in the two quadratures and also
identical, i.e., quantified by the same variance. Finally, note that we are
using $[\hat{Q},\hat{P}]=2i$, so that the variance of the vacuum noise is
equal to $1$.

\bibitem{Weedbrook2004} C.~Weedbrook, A.~M.~Lance, W.~P.~Bowen, T.~Symul, T.~C.~Ralph, P.~K.~Lam, Phys. Rev. Lett. \textbf{93}, 170504 (2004).

\bibitem {Grosshans2002}F.~Grosshans and P.~Grangier, Phys. Rev. Lett.
\textbf{88}, 057902 (2002).

\bibitem {Grosshans2003}F. Grosshans \textit{et al.}, Nature \textbf{421},
238-241 (2003).

\bibitem {Nav06}M.~Navascu$\mathrm{\acute{e}}$s, F.~Grosshans, and
A.~Ac\'{\i}n, Phys.~Rev.~Lett. \textbf{97}, 190502 (2006).

\bibitem {Gar06}R.~Garc$\mathrm{\acute{\i}}$a-Patr$\mathrm{\acute{o}}$n and
N.~J.~Cerf, Phys.~Rev.~Lett. \textbf{97}, 190503 (2006).

\bibitem {Ren09}R.~Renner and J.~I.~Cirac, Phys.~Rev.~Lett. \textbf{102},
110504 (2009).

\bibitem {Pir08}S.~Pirandola, S. L. Braunstein, and S. Lloyd, Phys.~Rev.~Lett.
\textbf{101}, 200504 (2008).

\bibitem {Gross03}F.~Grosshans, N. J. Cerf, J. Wenger, R. Tualle-Brouri, and
P. Grangier, Quantum. Inf. Comput. \textbf{3}, 535--552 (2003).

\bibitem {Kok2010}P.~Kok and B.~Lovett, \textit{Introduction to optical
quantum information processing} (Cambridge University Press, Cambridge, 2010).

\bibitem {Hol73}A.~S.~Holevo, Probl. Inf. Transm. \textbf{9}, 177-183 (1973).

\bibitem {Notation2}In DR one party guesses the \textit{encoding} of the
other, while in RR it is the \textit{decoding} to be guessed. In the two-way
protocol, the encoding is Alice's random displacement $a$, while the decoding
is Bob's post-processed variable $b$.

\bibitem {Sha48}C.~E.~Shannon, Bell Syst. Tech. J. \textbf{27}, 623-656 (1948).

\bibitem {Hol99}A.~S.~Holevo, M.~Sohma, and O.~Hirota, Phys. Rev. A
\textbf{59}, 1820-1828 (1999).



\bibitem {Eisert2002}J.~Eisert, S.~Scheel, and M.~B.~Plenio, Phys. Rev. Lett.
\textbf{89}, 137903 (2002).

\bibitem {Fiurasek2002}J.~Fiur\`{a}\v{s}ek, Phys.~Rev.~Lett. \textbf{89},
137904 (2002).

\bibitem {Gae}G. Spedalieri, C. Ottaviani, and S. Pirandola, Open Syst. Inf.
Dyn. \textbf{20}, 1350011 (2013).

\bibitem {Ger05}C.~C.~Gerry and P.~L.~Knight, \textit{Introductory Quantum
Optics} (Cambridge University Press, Cambridge, 2005).

\bibitem {geminiIR}Data on the transmission of infrared radiation in the
atmosphere is obtained from http://www.gemini.edu/?q=node/10789.


%\bibitem {2wayPROOF}C. Ottaviani and S. Pirandola, in preparation.
\end{thebibliography}
\end{document}